\documentclass[10pt,journal, letterpaper]{IEEEtran}

\usepackage{graphics}
\usepackage{graphicx}
\usepackage{amsmath}
\usepackage{amsthm}
\usepackage{enumerate}
\usepackage{amsfonts}
\usepackage{textcomp}
\usepackage{pstricks}
\usepackage{mathtools}
\usepackage[nolist]{acronym}

\newcommand{\figsize}{0.925 \columnwidth}

\begin{acronym}
    \acro{WLAN}{Wireless Local Area Network}
    \acro{RAT}{Radio Access Technology}
    \acro{RL} {Reinforcement Learning}
    \acro{PGRL} {Policy Gradient Reinforcement Learning}
    \acro{OFDMA}{Orthogonal Frequency-Division Multiple Access}
    \acro{NGMN} {Next Generation Mobile Network}
    \acro{WiMAX}{Worldwide Interoperability for Microwave Access}
    \acro{LTE}{Long Term Evolution}
    \acro{RAN}{Radio Access Networks}
    \acro{QoS}{Quality of Service}
    \acro{ICIC}{Inter-Cell Interference Coordination}
    \acro{PC}{Power Control}
    \acro{FFR}{Fractional Frequency Reuse}
    \acro{FL}{Fractional Load}
    \acro{BCR}{Block Call Rate}
    \acro{eNB}{enhanced eNode B}
    \acro{CDMA}{Code Division Multiple Access}
    \acro{GSM}{Global System for Mobile Communications}
    \acro{BS}{Base Station}
    \acro{PRB}{Physical Resource Block}
    \acro{NG}{Next Generation}
    \acro{B3G}{Beyond 3G}
    \acro{SON}{Self-Organizing Networks}
    \acro{3GPP}{3rd Generation Partnership Project}
    \acro{UE}{User Equipment}
    \acro{RRM}{Radio Resource Management}
    \acro{PS}{Packet Scheduling}
    \acro{KPIs}{Key Performance Indicators}
    \acro{SINR}{Signal to Interference plus Noise Ratio}
    \acro{HSDPA}{High Speed Downlink Packet Access}
    \acro{RR}{Round Robin}
    \acro{TDMA}{Time Division Multiple Access}
    \acro{MTP}{Max Throughput}
    \acro{PF}{Proportional Fair}
    \acro{MMF}{Max-Min Fair}
    \acro{MAB}{Multi-Armed-Bandit}
    \acro{RB}{Restless Bandit}
    \acro{ODE}{Ordinary Differential Equation}
    \acro{i.i.d}{independent and identically distributed}
    \acro{MIMO}{Multiple Input Multiple Output}
    \acro{c.d.f}{cumulative distribution function}
    \acro{p.d.f}{probability density function}
    \acro{PPRBS}{Per Physical Resource Block Scheduler}
    \acro{KPI}{Key Performance Indicator}
    \acro{SIC}{Successive Interference Cancellation}
    \acro{FTP}{File Transfer Protocol}
	\acro{PSO}{Particle Swarm Optimization}
    \acro{FTT}{File Transfer Time}
    \acro{BS}{Base Station}
    \acro{a.s}{almost surely}
    \acro{AWGN}{Additive White Gaussian Noise}
    \acro{CSI}{Channel State Information}
    \acro{MDP}{Markov Decision Process}
    \acro{4G}{4th Generation}

\end{acronym}

\newtheorem{theorem}{Theorem}

\newtheorem{corollary}{Corollary}

\newcommand{\RR}{\mathbb{R}}
\newcommand{\NN}{\mathbb{N}}
\newcommand{\PP}{\mathbb{P}}
\newcommand{\AAb}{\mathbb{A}}
\newcommand{\II}{\mathbb{I}}
\newcommand{\tends}[2]{\underset{#1 \to #2}{\to}}
\newcommand{\limu}[2]{ \underset{#1 \to #2} {\lim} \,}

\DeclarePairedDelimiter\Set{\{}{\}}

\newcommand{\expec}[1]{E \left[ #1 \right] }

\newcommand{\argmin}[1]{ \underset{#1} {\arg \min} \,}
\newcommand{\argmax}[1]{ \underset{#1} {\arg \max} \,}

\newcommand{\maxu}[1]{ \underset{#1} {\max} \,}

\newcommand{\limsupu}[1]{ \underset{#1} {\limsup} \,}

\newcommand{\indic}{{\bf 1} }

\begin{document}

\title{The association problem in wireless networks: a Policy Gradient Reinforcement Learning approach}
\author{

\IEEEauthorblockN{Richard Combes\IEEEauthorrefmark{1}, Ilham El Bouloumi\IEEEauthorrefmark{2}\IEEEauthorrefmark{3}, Stephane Senecal\IEEEauthorrefmark{2}, Zwi Altman\IEEEauthorrefmark{2}}

\IEEEauthorblockA{\IEEEauthorrefmark{1} KTH, Royal Institute of Technology\\
Stockholm, Sweden\\
rcombes@kth.se}

\IEEEauthorblockA{\IEEEauthorrefmark{2}France Telecom R\&D/Orange Labs\\
38-40 rue du G\'en\'eral Leclerc\\
92794 Issy-les-Moulineaux CEDEX 9, France\\
\{zwi.altman, stephane.senecal\}@orange.com}

\IEEEauthorblockA{\IEEEauthorrefmark{3}Telecom Bretagne\\
Technopole Brest-Iroise - CS 83818\\
29238 Brest CEDEX 3, France\\
ilham.elbouloumi@telecom-bretagne.eu}
}

\maketitle

\begin{abstract}
	The purpose of this paper is to develop a self-optimized association algorithm based on \ac{PGRL}, which is both scalable, stable and robust. The term robust means that performance degradation in the learning phase should be forbidden or limited to predefined thresholds. The algorithm is model-free (as opposed to Value Iteration) and robust (as opposed to Q-Learning). The association problem is modeled as a Markov Decision Process (MDP). The policy space is parameterized. The parameterized family of policies is then used as expert knowledge for the \ac{PGRL}. The \ac{PGRL} converges towards a local optimum and the average cost decreases monotonically during the learning process. The properties of the solution make it a good candidate for practical implementation. Furthermore, the robustness property allows to use the \ac{PGRL} algorithm in an ``always-on" learning mode.
	\footnote{This work has been partially supported by the Agence Nationale de la Recherche within the project ANR-09-VERS0: ECOSCELLS.}
\end{abstract}

\section{Introduction}\label{sec:Introduction}
The association problem in wireless networks has received considerable interest in the past few years due to its various applications. First, the mobile network landscape has become more and more heterogeneous. The network operator often needs to manage different radio access technologies such as \ac{GSM}, \ac{HSDPA}, \ac{LTE}, and \ac{WLAN}. In this context, connecting to a network with lower load by means of advanced \ac{RRM} algorithms, via both mobility and selection/re-selection mechanisms, can significantly impact network performance and perceived QoS (see for example \cite{SalahCooperativeHO}). The complexity of managing resources in highly heterogeneous networks has been one of the drivers for the new paradigm of partial shift of the resource management burden from the network to the user terminals which can learn how to take intelligent association decisions (\cite{P1900.4,Majed_CR_Association}).

The renewed interest in the association problem has appeared with the introduction of \ac{SON} in \ac{4G} mobile networks. \ac{SON} covers self-configuration, self-optimization and self-healing (automatic troubleshooting).  In \ac{LTE}, \ac{SON} has already been introduced in the first Release (Release 8) of the standard \cite{3gpp.36.300}. The intra-system mobility load balancing optimization, which is closely related to the association problem, is one of the first self-optimization features introduced in the \ac{LTE} standard \cite{3gpp.36.902}.

Self-optimization aims at adapting the network to traffic variations and to new conditions of operation. The self-optimization process can be performed by autonomously adjusting network parameters such as parameters of \ac{RRM} algorithms. The adoption of self-optimizing functionalities in real operating networks introduces strict requirements such as scalability and stability. Scalability means that the \ac{SON} features should operate correctly when deployed in many network nodes, such as base stations and their neighboring ones. Stability means that the network empowered by the SON functionality diminishes congestion in the network so that the number of active users remains bounded and tends to a stationary regime. This stability definition corresponds to the stability in queuing systems.

Deriving optimal parameters or controllers via a learning process such as \ac{RL} (\cite{SuttonBarto}) often requires a learning (or exploration) phase. Monotonic performance improvement during the learning phase is sought. We call this property \emph{robust learning}.

The purpose of this paper is to develop a self-optimized association algorithm based on \ac{PGRL}, which is both scalable, stable and robust. The requirement of robustness excludes direct application of \ac{RL} solutions such as Q-Learning (\cite{SuttonBarto}). Value Iteration (\cite{SuttonBarto}) does not apply either since we assume no knowledge of the system dynamics, namely  the transition probabilities of the \ac{MDP}. The association problem is modeled as a \ac{MDP}, (cf. \cite{PutermanMDP,SuttonBarto}), and its optimal policy is derived for a small size problem to learn a functional form and to parameterize the policy space. Then, the obtained solution is used as \emph{expert knowledge} for the \ac{PGRL} (\cite{WilliamsReinforce,BaxterBartlettPolicyGradient1,BaxterBartlettPolicyGradient2}). The \ac{PGRL} converges to a local optimum and the average cost decreases monotonically during the learning phase. This makes it a good candidate for practical implementation. Furthermore, the robustness property allows to use the \ac{PGRL} algorithm in a \emph{``always on" learning} mode.\\
The contributions of the present paper are the following:
\begin{enumerate}[(i)]
\item A queuing model for the problem of association in wireless networks is stated. This model takes into account flow-level dynamics allowing to optimize end-to-end, user-level performance indicators such as network capacity or mean file transfer time.
\item It is shown that the static association problem is tractable by classical convex optimization techniques.
\item The dynamic association problem is modeled as a \ac{MDP}. A reinforcement learning (on-line learning) solution is proposed, and its convergence to a local optimum is proven. The approach is scalable when the number of \ac{BS} increases, e\-na\-bling to develop a practical solution.
\item A heuristic scheme is proposed which allows the algorithm to operate in a fully distributed manner and to greatly improve the accuracy of the gradient estimates.
\end{enumerate}
The paper is organized as follows. Section~\ref{sec:System model and problem statement} describes the system model for a wireless network serving elastic traffic, taking into account flow-level dynamics, and states the association problem. Section~\ref{sec:The static association problem} examines the static version of the association problem, and shows that the problem is tractable by classical convex optimization techniques. Section~\ref{sec:The dynamic association problem} presents the dynamic case, and models it as a \ac{MDP}. A family of parameterized policies is introduced, allowing to develop a scalable reinforcement learning approach when the number of \acp{BS} grows. A heuristic which allows the algorithm to operate in a fully distributed manner and to improve the accuracy of the gradient estimation is introduced. Section~\ref{sec:Numerical Experiments} presents numerical experiments showing that the proposed method effectively increases the network capacity, and that the proposed heuristic considerably improves the accuracy and convergence speed of the method. Section~\ref{sec:Conclusion} concludes the paper.

\section{System model and problem statement}\label{sec:System model and problem statement}
We describe here the system model encompassing the PHY, MAC and application layers. For each layer, we summarize relevant results in the literature and state the association problem. We consider the downlink of a wireless network, serving elastic traffic. The system bandwidth is $W$, under full reuse. The network area is $\AAb \subset \RR^2$, and we assume it to be bounded. We denote by $N_s$ the number of \acp{BS}.
	\subsection{Physical layer}
	Consider a single user located at $r \in \AAb$, served by \ac{BS} $s$ with $1 \leq s \leq N_s$ . We write $S_s(r)$ its \ac{SINR}. We consider \ac{AWGN} and channel fading. We assume that the fading process is ergodic. We assume block fading i.e the fading process remains constant for the duration of a codeword. We treat the interference as Gaussian noise. We consider that full \ac{CSI} is available at the receiver. Given $Z \in \mathbb{C}$ a value of the fading process, the data rate of a user is given by a certain function $\phi$:
\begin{equation}
	\phi(|Z|^2 S_s(r)) \leq W\log_2(1 + |Z|^2 S_s(r))
	\label{eq:datarateuser}
\end{equation}
For a large number of codewords, the time average of the the data rate of a user (\ref{eq:datarateuser}) is the ergodic data rate:
\begin{equation}
R_s(r) = \expec{\phi(|Z|^2 S_s(r))}
\label{eq:Rsr}
\end{equation}
\subsection{MAC layer}	
$n_s$ users are served simultaneously by BS $s$. For non-opportunistic scheduling, all users receive an equal part of the radio resources, and the throughput of a user located at $r$ is equal to $\frac{R_s(r)}{n_s}$. This corresponds to Round-Robin scheduling. In the case of opportunistic scheduling, each user is allocated the channel when its fading is the best. Define $\Set{Z_i}_{1 \leq i \leq n_s}$ \ac{i.i.d} copies of $Z$, the throughput of a user located at $r$ is equal to:
\begin{equation}
\expec{ (\prod_{i=2}^{n_s} \indic_{\Set{|Z_i| \leq |Z_1|}}) \phi(|Z_1|^2 S_s(r))}
\label{eq:datarateuser2}
\end{equation}
We approximate this quantity by $\frac{g(n_s) R_s(r)}{n_s}$. The function $g$ is non-decreasing and denotes the multi-user diversity gain, and where $g(n_s) \geq 1$. We write $g(\infty) = \limu{n_s}{+\infty} g(n_s)$ the maximal diversity gain. The non-opportunistic scheduling case can be seen as a particular case with $g \equiv 1$. Derivation of $g$ for particular channel models can be found in \cite{CombesPEVA2011,CombesCoverageCapacityMIMO2011,CombesWiopt2011}.
	\subsection{Application layer}
Consider users arriving randomly according to a spatial Poisson process on $\AAb \times \RR^{+}$, with intensity $\lambda(dr \times dt) = \lambda dr dt$. We write $\lambda_{tot} = \lambda \int_{\AAb} dr$ the total arrival rate. The arrival process is marked with $\sigma$, the file size to be downloaded and we assume independence between the arrival process and the file sizes. We write $\AAb_s \subset \AAb$ the area served by \ac{BS} $s$. We say that the system is stable if the distribution of the number of active users tends to a stationary limit, and unstable if the number of active users grows to infinity. Such a system can be modeled by $N_s$ parallel M/G/1/PS (Processor Sharing) queues, and the following theorem summarizes the results on the system stability region and the mean performance (cf. \cite{Bonald-dimensioning}).
	
\begin{theorem}\label{thm:MG1PS}
The load of \ac{BS} $s$ is
\begin{equation}\rho_s = \frac{\lambda \expec{\sigma}}{g(\infty)} \int_{\AAb_s} \frac{1}{R_s(r)}dr, \end{equation}
\ac{BS} $s$ is stable if $\rho_s < 1$ and unstable if $\rho_s > 1$.\\
Consider non opportunistic scheduling and assume stability of \ac{BS} $s$, and denote by $n_s(t)$ the mean number of active users served by \ac{BS} $s$ in stationary regime at time $t$. Then we have that: \begin{equation}\expec{n_s(t)} = \frac{\rho_s}{1 - \rho_s} \end{equation}
\end{theorem}
	
\subsection{The association problem}
Now let us consider a new zone $\AAb_0 \subset \AAb$. The association problem consists in allocating the traffic arriving in $\AAb_0$ to \acp{BS}, in order to optimize a given performance indicator. We distinguish two problems: the static association problem and the dynamic association problem.
	
In the static version, the traffic is attributed to \acp{BS} regardless of the current state of the system. Namely, $\AAb_0$ is partitioned into $N_s$, regions, and users arriving in the $s$-th region will be served by \ac{BS} $s$ regardless of the number and locations of active users.
In the dynamic problem, the system has access to the current user configuration to make a decision. Namely, the user configuration is composed of the number of active users, their locations, the \ac{BS} they are currently attached to and their remaining amount of data to be downloaded. We call a policy a mapping between user configuration and association of each user. The problem consists in finding the policy that maximizes a given performance indicator.
	
	\subsection{A finite set of data rates}
	In practical systems, there only exists a finite set of possible data rates, due to the finite number of modulation and coding schemes. This allows us to introduce a discretized version of the previous model, used in next sections.  Let ${\cal I} \in \NN$ denote the number of possible data rates, and $R^{(i)}$ - the $i$-th possible data rate. We use the convention $R^{(0)} = 0$ and $R^{({\cal I}+1)} = +\infty$. We write 
	\begin{equation}
	\AAb_{s,i} = \Set{r \in \AAb_s | R_s(r) \in  [R^{(i)},R^{(i+1)})}.
	\end{equation} 
	We assume that all users in $\AAb_{s,i}$ served by \ac{BS} $s$ have a data rate of $R^{(i)}$. 

	We assume that $\AAb_{s,0}$ is empty, which can be enforced through admission control, namely, any user whose radio condition does not enable him to achieve even the lowest allowed data rate is not allowed to enter the system. $\AAb_s$ is partitioned into $\cup_{1 \leq i \leq {\cal I}} \AAb_{s,i}$. 
	
	 It is noted that the discretized model is a conservative model, since user data rates in the discretized model are lower bounds for the data rates in the continuous model. This is an important property since it implies that system performance given by the discrete model is a lower bound of the system performance in the continuous model, and instability in the continuous model implies instability in the discrete model as well.
	
	We assume that $r \mapsto R_s(r)$ is measurable for all $s$, hence $\AAb_{s,i}$ are Borel sets for all $(s,i)$, and the integrals for the system performance in Theorem~\ref{thm:MG1PS} are well-defined. We call users arriving in $\AAb_{s,i}$ users of class $(s,i)$. Their arrival rate is $\lambda_{s,i} = \lambda \int_{\AAb_{s,i}} dr$.
	
	We partition $\AAb_0$ as well. We write  $\II = \Set{0 ,\cdots , {\cal I}}^{N_s} \setminus \Set{(0,\dots,0)}$, we consider $I \in \II$ and denote by $I_s$ its $s$-th component. We define the zone associated to configuration $I$, $\AAb_{0,I}$, by:
\begin{equation}
	\AAb_{0,I} = \Set{r \in \AAb_{0} | R_s(r) \in  [R^{(I_s)},R^{(I_s+1)}) , 1 \leq s \leq N_s}.
\end{equation}
	We have that $\AAb_0 = \cup_{I \in \II} \AAb_{0,I}$.  We denote users of class $(0,i)$ users that have arrived in $\AAb_{0,I}$. Their arrival rate is $\lambda_{0,I} = \lambda \int_{\AAb_{0,I}} dr$.

\begin{figure}[htpb]
		\begin{center}
			\includegraphics[width=\columnwidth]{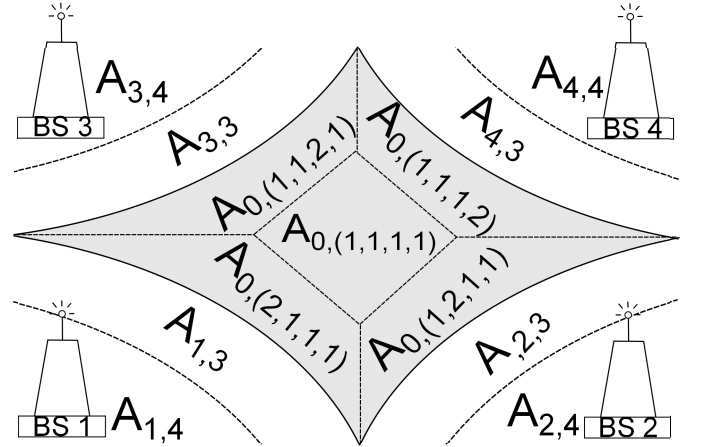}
		\end{center}
		\caption{The association problem}
		\label{fig:association_problem}
	\end{figure}

	Figure \ref{fig:association_problem} represents the model with $4$ possible data rates and $4$ \acp{BS}. The gray zones belong to $\AAb_{0}$ and can be served by any \ac{BS}. The non-gray zones can only be served by the closest \ac{BS}. Zone $\AAb_{1,4}$ is the closest to \ac{BS} $1$ and can only be served by \ac{BS} $1$ with data rate $R^{(4)}$. Zone $\AAb_{1,3}$ can only be served by \ac{BS} $1$, but with lower data rate $R^{(3)}$ since it is further away from \ac{BS} $1$. The gray zone $\AAb_{0,(2,1,1,1)}$ can be served by $\ac{BS}$ $1$ with data rate $R^{(2)}$ and by other \acp{BS} with data rate $R^{(1)}$, since it is closer to \ac{BS} $1$. The central zone $\AAb_{0,(1,1,1,1)}$ can be served by all \acp{BS} with data rate $R^{(1)}$.
	
	The association problem is to determine the proportion of traffic from $\AAb_{0,I}$ to be served by \ac{BS} $s$, for all $s$ and $I$. 
	
\subsection{Numerical tractability}	
	
For each zone $\AAb_{0,I}$, $I \in \II$, we need to specify the amount of traffic which will be associated to each \ac{BS}. There are $({\cal I}+1)^{N_s} - 1$ such zones, and the number of variables needed for the association problem is $N_s ( ({\cal I}+1)^{N_s} - 1)$. The number of variables grows exponentially with $N_s$ and makes the problem numerically intractable, simply from the memory required to store a solution. In practice, however, the vast majority of those zones will be empty. Namely, for a given zone $\AAb_{0,I}$, if there exists a \ac{BS} $s$ such that the available data rate at station $s$ is high, then all the users arriving in $\AAb_{0,I}$ will be allocated to \ac{BS} $s$ and generate little load. Hence $\AAb_0$ typically consists of cell edges zones. We will assume that there exists ${\cal I}_{edge} < {\cal I}$ such that, if users arriving in $\AAb_{0,I}$ can be served by a station $s$ at a rate larger than $R^{({\cal I}_{edge})}$ i.e  $R^{(I_s)} \geq R^{({\cal I}_{edge})}$, then they will all be served by \ac{BS} $s$.
	
	Furthermore, consider a location $r$, which is far away from the location of \ac{BS} $s$. Then we will have $R_s(r) < R^{(1)}$ which means that no traffic from $\AAb_{0,I}$ shall be allocated to $s$ anyway. We say that $(s,r)$ is connected if $R_s(r) \geq R^{(1)}$ and we say that \acp{BS} $s$ and $s^{\prime}$ are neighbors if there exists a point $r$ such that $(s,r)$ and $(s^{\prime},r)$ are connected. We write $N_{neigh}$ the maximal number of neighbors for a \ac{BS}. The actual number of variables can then be crudely upper bounded by $N_s N_{neigh} ({\cal I}_{edge}^{N_{neigh}} - 1)$ which only grows linearly $N_s$. Typical values are $N_{neigh} = 6$ and ${\cal I}_{edge} = 1$, for which we need $6 N_s$ variables, and a large number of \acp{BS} can be easily handled.
	
\section{The static association problem}\label{sec:The static association problem}
	We consider non-opportunistic scheduling. We write $a_{s,I} \in [0,1]$ the proportion of users of class $(0,I)$ served by \ac{BS} $s$. This can correspond to three different physical mechanisms, each of which might or might not be applicable depending on the network technology:
	\begin{enumerate}[(i)]
	\item A user arriving in $\AAb_{0,I}$ attaches itself to \ac{BS} $s$ with probability $a_{s,I}$.
	\item A user arriving in $\AAb_{0,I}$ downloads a fraction $a_{s,I}$ of the requested file through \ac{BS} $s$. 
	\item $\AAb_{0,I}$ is divided into $N_s$ sub-regions $\AAb_{0,I,s}$ the sizes of which are proportional to $a_{s,I}$. Namely: \begin{equation}a_{s,I}=  \frac{\int_{\AAb_{0,I,s}} dr}{\int_{\AAb_{0,I}} dr} \end{equation}
	\end{enumerate}
We define ${\bf a} = (a_{s,I})_{1 \leq s \leq N_s , I \in \II}$. Given ${\bf a}$, the load of \acp{BS} can be calculated as in Theorem~\ref{thm:MG1PS}:
\begin{equation}
\rho_s({\bf a}) = \expec{\sigma} \left(  \lambda  \int_{\AAb_s} \frac{1}{R_s(r)}dr  +  \sum_{I \in \II} \frac{a_{s,I} \lambda_{s,I}}{R^{(I_s)}} \right)
\label{loads_optim}
\end{equation}
We write ${\bf \rho}({\bf a}) =  (\rho_1({\bf a}),\cdots,\rho_{N_s}({\bf a}))$.\\
Let us consider a convex function $U:\RR^{N_s \times | \II |} \to \RR$, the static association problem corresponds to the following optimization problem:
	\begin{align} &\text{minimize  } &U({\bf \rho}({\bf a})) \label{optimization problem}\\
	&\text{subject to} &a_{s,I} \geq 0 &\;,\; 1 \leq s \leq N_s &, I \in \II  \label{optimization constraint 1}\\
	&\text{and} & \sum_{1 \leq s \leq N_s} a_{s,I} = 1 &, I \in \II \label{optimization constraint 2}
	\end{align}
	Such optimization problems bear strong resemblance with optimization problems encountered in routing.
	\begin{theorem}\label{th:optim}
Optimization problem~(\ref{optimization problem}) is convex.
\end{theorem}
\begin{proof}
As seen in equation~\eqref{loads_optim}, ${\bf a} \mapsto {\bf \rho}({\bf a})$ is affine, hence by composition of an affine mapping with a convex function, we have that ${\bf a} \mapsto U({\bf \rho}({\bf a}))$ is convex. Similarly, constraints~\eqref{optimization constraint 1} and~\eqref{optimization constraint 2} are affine hence convex.
	\end{proof}
	\begin{corollary}
		The minimization of mean file transfer time is a particular case of \eqref{optimization problem}, with
		\begin{equation} U({\bf \rho}({\bf a})) = \frac{1}{\lambda_{tot}} \sum_{1 \leq s \leq N_s} \frac{\rho_s({\bf a})}{1 - \rho_s({\bf a})} \end{equation}
	\end{corollary}
\begin{proof}
By Little's law~\cite{little}, the mean file transfer time is equal to the mean number of active users divided by the total arrival rate. From Theorem~\ref{thm:MG1PS}, the mean file transfer time is
\begin{equation}
	\frac{1}{\lambda_{tot}} \expec{\sum_{1 \leq s \leq N_s} n_s(t)} = \frac{1}{\lambda_{tot}} \sum_{1 \leq s \leq N_s} \frac{\rho_s}{1 - \rho_s}
	\label{eq:mftt}
\end{equation}
Furthermore, let $x \in \RR^{N_s}$, we have that
\begin{equation}
x \mapsto \frac{1}{\lambda_{tot}}.\frac{x_s}{1 - x_s}=   \frac{1}{\lambda_{tot}}\left(\frac{1}{1 - x_s} - 1\right)
\label{eq:func1}
\end{equation}
which is convex and
\begin{equation}
x \mapsto \frac{1}{\lambda_{tot}}\sum_{1 \leq s \leq N_s} \frac{x_s}{1 - x_s}
\label{eq:func2}
\end{equation}
is convex as a sum of convex functions.
	\end{proof}
	Theorem \ref{th:optim} shows that~(\ref{optimization problem}) is computationally tractable using classical convex optimization methods (see for example \cite{ConvexOptim}), and it includes the important problem of minimizing the mean file transfer time. The reader can refer to the literature on routing in which such optimization problems have been studied extensively.

\section{The dynamic association problem}\label{sec:The dynamic association problem}
\subsection{MDPs and reinforcement learning}
\label{sec:rl}
	\acp{MDP} can be used to model optimal control problems where the system state has a markovian structure as described briefly below. The reader can refer to \cite{PutermanMDP,SuttonBarto} for a complete exposition of the topic.
	
	 We consider $(\Omega,\PP,{\cal F})$ a probability space. A \ac{MDP} is defined by a discrete set of possible system states ${\cal S}$, a discrete set actions ${\cal A}$, an intensity matrix $\mu:{\cal S}^2 \times {\cal A} \to \RR^+$ , and a cost function $r:{\cal S} \times {\cal A} \times \Omega \to \RR$. 
	 
	  Denote $R(s,a) = \expec{r(s,a,\omega)}$ the average cost. A policy is a mapping ${\cal S} \mapsto {\cal D}({\cal A})$, where ${\cal D}({\cal A})$ is the space of probability distributions on ${\cal A}$. Given a policy $P$ and $P(s,a)$ the probability of choosing action $a$ in state $s$ we write $\mu_P(s^{\prime},s) = \sum_{a \in {\cal A}} P(s,a) \mu(s^{\prime},s,a)$ the transition intensity from $s$ to $s^{\prime}$ under policy $P$.
	  
 Consider a policy $P$, we call the stochastic process $\Set{s_t,a_t,r_t}$ a realization of the \ac{MDP} for policy $P$ if: 
	
\begin{itemize}	
	\item $a_{t}$ depends only on $s_t$ and is distributed according to $P(s_t)$
	\item $\Set{s_{t}}$ is a Markov process with intensity matrix $\mu_P(s^{\prime},s)$
	\item $r_t$ depends only on $(s_{t},a_{t})$, and is equal to $r(s_{t},a_{t})$ in distribution	
\end{itemize}

	The average cost of policy $P$ starting at $s \in {\cal S}$ can be defined by:
	
	\begin{equation} J_{P}(s) = \limsupu{T \to +\infty} \frac{1}{T} \expec{\int_{0}^{T} r_t dt}. \end{equation}

	Assume that $P$ is such that $\Set{s_t}$ is ergodic, then $s \mapsto J_{P}(s)$ is constant and we write $J_{P}(s) = J_{P}$. In the following, we will use the hypothesis that the \ac{MDP} is such that either $P$ makes $\Set{s_t}$ ergodic, or else $J_{P}(s) = +\infty$. While this might first appear as a strong hypothesis, this is actually the case in a large number of problems such as queuing problems. Namely, either a policy makes the queue ergodic, or else the number of active users grows to infinity.

	Solving the \ac{MDP} consists then in finding the optimal policy which minimizes the cost $P^* = \argmin{P} J_{P}$. Existence and uniqueness results can be found in \cite{PutermanMDP,Blackwell}.

	The reinforcement learning problem is defined as deriving the optimal policy without the knowledge of the probabilistic structure of the model, through trial-and-error, cf. \cite{SuttonBarto}. Namely, the intensity matrix $\mu$ and the distribution of the costs $r$ are unknown, and we can only obtain \ac{MDP} realizations $\Set{s_t,a_t,r_t}$. It is hence a simulation-based method or model-free method. 

\subsection{Resolution techniques and scalability}

	It is noted that by using discretization and ``uniformization'' (cf. \cite{PutermanMDP}), we can always reduce the continuous time \ac{MDP} to a discrete time \ac{MDP}. In the remainder of the paper, whenever we employ reinforcement learning, it will be done using the discrete time version of the system.

	When the intensity matrix and the distribution of the costs are known, and both ${\cal S}$ and ${\cal A}$ are is finite, the optimal policy can be derived via an iterative scheme, namely dynamic programming thanks to a fixed-point relation holding at optimality. In practice however, for large state spaces, this becomes numerically intractable (``curse of dimensionality''). 

	A scalable approach is to introduce a parameterized family of policies $\Set{P_{\theta}, \theta \in \Theta}$ and define the cost $J(\theta) = J_{P_{\theta}}$, once again using the previous hypothesis of ergodicity. This para\-me\-te\-rization is a powerful idea for solving optimal control problems numerically with state spaces of large dimension. The problem becomes the optimization of $\theta \mapsto J(\theta)$, which is assumed to be computationally tractable. It is noted that the performance of such a scheme highly depends on the goodness of the chosen family of policies. Choosing a good parametrization generally implies having some knowledge on the structure of the optimal controller. It can be seen as a form of ``expert knowledge''.

	\subsection{Policy gradient reinforcement learning approach}
	
	It remains to show how to optimize $\theta \mapsto J(\theta)$, without the knowledge of the probabilistic structure of the model. 
 We assume that we are at least able to simulate the system for a fixed value of $\theta$, and that $J(\theta)$ can then be computed by averaging the observed cost for a sufficiently long simulation.\\
 We are interested in both local and global optima.
For global optima, since the problem is not convex in general, a search heuristic (e.g. genetic algorithm, particle swarm optimization) is needed, which requires to compute $J(\theta)$ for a large number of values of  $\theta$.\\
For local optima, we can use a descent method by calculating  $\nabla_{\theta} J(\theta)$. The crudest approach is to approximate the gradient using finite differences. We compute its $k$-th component by:
	\begin{equation}\frac{J(\theta + \epsilon e_k) - J(\theta - \epsilon e_k)}{2 \epsilon},\end{equation}
for a small $\epsilon$, where $e_k$ stands for the $k$-th unit vector. This is possible whenever $J(\theta)$ can be computed, but requires a number of simulations equal to twice the number of components of $\theta$. Furthermore, this approach is not suitable for an ``on-line'' implementation where instead of simulating the system, the algorithm must compute an estimate of the gradient based on observations from a real network.\\
	The approach we propose here consists in estimating $\nabla_{\theta} J(\theta)$ from a single (discrete-time) sample path $\Set{s_t,a_t,r_t}_{t \in \NN}$, using the method described in \cite{BaxterBartlettPolicyGradient1}. 

 We write $\PP_{\theta}$ the probability measure given policy $P_{\theta}$. It is possible to compute iteratively the eligibility traces:
\begin{equation}
	z(t+1)=\beta z(t) + \frac{\nabla_\theta \PP_{\theta}(a_t | s_t)}{\PP_{\theta}(a_t | s_t)},
	\label{eq:zt}
\end{equation}
and the associated gradient estimates:
\begin{equation}
	\Delta(t+1)=\Delta(t) + \frac{1}{t+1}(r_{t+1}z(t+1)-\Delta(t)).
	\label{eq:Deltat}
\end{equation}
The gradient estimates converge \ac{a.s} to an ascent direction:
\begin{equation}
	\liminf_{t \rightarrow + \infty} < \Delta(t) ,\nabla_{\theta} J(\theta) > \,\,> 0.
	\label{eq:convnabeta}
\end{equation}

\subsection{Modeling the association problem as a MDP}
\label{sec:mdp}

	We show that the association problem can be modeled as a continuous time \ac{MDP}. We assume that association decisions are taken when users enter the system. This avoids the possibility of constant hand-overs every time the user configuration changes which would be impractical due to a high amount of additional overhead. We need to introduce several artificial states in order to use a \ac{MDP} model.
	
	We write $n_{s,I}$ the number of users of class $(0,I)$ that are attached to \ac{BS} $s$. The user configuration is ${\bf n} = ( n_{s,I} , n_{s,i})_{1 \leq s \leq N_s, I \in \II}$ which completely specifies the number of users of each class, and the \ac{BS} they are attached to.
	 
	 When the user configuration is ${\bf n}$ and a user of class $(0,I)$ arrives in the network, the system enters an ``artificial'' state denoted $({\bf n},I)$. We assume that the time spent in $({\bf n},I)$ is $0$ and transition to $( {\bf n}^{\prime},0)$ is instantaneous, where ${\bf n}^{\prime}$ is the new user configuration, depending on the association decision taken by the system. We will denote states of the type $({\bf n},I)$ as ``arrival states'' and states of the type $({\bf n},0)$ as ``ordinary states''.
	 
	In ordinary states, no action is available to the controller. For arrival state $({\bf n},I)$, the available actions are the \acp{BS} to which users of class $(0,I)$ can be attached. It is noted that the subset of states in which an action is available is relatively small, which is attractive in terms of practical controller implementation.
	 
	 We choose the cost of a state as the total number of users in this state. Using ergodicity of policies, and Little's Law~\cite{little}, we have that $J_{P}$ divided by the total arrival rate is in fact the mean file transfer time under the policy $P$. Alternatively, we can define the cost to be $1$ if at least one user has a throughput smaller than a target data rate, and $0$ otherwise. $J_{P}$ is then the outage probability under the policy $P$.
	 
	Assuming that file sizes are exponentially distributed, then the system is a continuous time \ac{MDP}, and we specify the transition intensities. There are five types of transitions, and we introduce a shorthand notation:
\begin{itemize}
\item Arrival of a user of class $(s,i)$ , $s > 0$ (denoted $(s,i)^{++}$)
\item Arrival of a user of class $(0,I)$ , (denoted $(0,I)^{++}$)
\item Attachment of a user of class $(0,I)$ to \ac{BS} $s$ (denoted $(0,I) \to s$)
\item Departure of a user of class $(s,i)$ , $s > 0$ (denoted $(s,i)^{--}$)
\item Departure of a user of class $(0,I)$ attached to \ac{BS} $s$ (denoted $(s,I)^{--}$)
\end{itemize}
	
	For user configuration ${\bf n}$, we write $T_s({\bf n})$ the number of users served by \ac{BS} $s$:
\begin{equation}
	T_s({\bf n}) = \sum_{I \in \II} n_{s,I} +  \sum_{i = 1}^{\cal I} n_{s,i}.
	\label{eq:ts}
\end{equation}
We also define the number of users served by \ac{BS} $s$ for which the peak rate is equal to $R^{(i)}$: 
\begin{equation}
	T_{s,i}({\bf n}) = n_{s,i} + \sum_{I \in \II} n_{s,I} {\bf 1}_{ \Set{i} }(I_s)  .
	\label{eq:tsi}
\end{equation}

	The transition intensities for arrivals can be written:
\begin{equation}
	\mu( (s,i)^{++}) = \lambda_{s,i} = \lambda \int_{\AAb_{s,i}} dr,
	 \label{eq:sipp1}
\end{equation}
and:
	 \begin{equation}
	\mu( (0,I)^{++}) = \lambda_{0,I} = \lambda \int_{\AAb_{0,I}} dr.
	 \label{eq:sipp2}
\end{equation}
	Let $\overline{\mu} > 0$ be a constant, the intensity for attachment of a user of class $(0,I)$ to \ac{BS} $s$ is:
\begin{equation}
	\mu( (0,I) \to (s,a)) = \left\{
	\begin{aligned}  
	\overline{\mu} \text{ if } a=s, \\
	0 \text{ otherwise.}
	 \end{aligned}
	 \right.
	 \label{eq:siar}
\end{equation}
The proportion of time spent in arrival states can be rendered arbitrarily small by setting $\overline{\mu}$ large enough. This allows to model a system in which the association decisions are instantaneous as specified previously. The departure intensities can be derived, for $s > 0$:
	\begin{equation}
	\mu( (s,i)^{--}) = \frac{1}{\expec{\sigma}} \frac{n_{s,i} R^{(i)}}{T_s({\bf n})},
	\label{eq:simm}
\end{equation}
	and for $s = 0$:
	\begin{equation}
	\mu( (s,I)^{--} ) = \frac{1}{\expec{\sigma}} \frac{n_{s,I} R^{(I_s)}}{T_s({\bf n})}.
	\label{eq:zimm}
\end{equation}
	It is noted that the only transitions for which the intensity actually depends on the chosen action are transitions linked to attachment of a user.
\subsection{Parameterization}

 Since all transition intensities have been specified, the optimal policy (minimizing the average cost) can be computed numerically, using Value Iteration (cf. \cite{SuttonBarto} for instance). However, this is only feasible when $N_s$ is small. Indeed, the size of ${\cal S}$ grows exponentially with $N_s$ (``curse of dimensionality''), and finding the optimal policy becomes intractable in practice.\\

 	Before stating the chosen parameterization, we give the rationale behind such a choice. Let $\theta \in \RR^{N_s \times ({\cal I}+1) \times |\II |}$ a vector of weights. When a user arrives in zone $\AAb_{0,I}$, he must evaluate, for each possible \ac{BS}, the peak rate available with this \ac{BS}, and the load of the \ac{BS}, which depends on the number of active users already attached to the \ac{BS} and their peak rates. In order to take a decision, for each \ac{BS} $s$, we compute the weighted sum $\theta_{s,0,I} +  \sum_{1 \leq i \leq {\cal I}} \theta_{s,i,I} T_{s,i}({\bf n})$. The term $\theta_{s,0,I}$  is independent of the load, and is linked to the peak rate available at \ac{BS} $s$, irrespective of its load.  The term $\sum_{1 \leq i \leq {\cal I}} \theta_{s,i,I} T_{s,i}({\bf n})$ is load dependent and is a weighted sum of the number of active users with different peak rates. $\theta_{s,i,I}$ is the weighting coefficient given to users attached to \ac{BS} $s$ with peak rate $R^{(i)}$. 
 
 Furthermore, the attachment rule must assign a positive probability to \emph{all} possible decisions, since we require the average cost to be differentiable with respect to $\theta$. When a user of class $(0,I)$ enters the network, he is attached to \ac{BS} $s$ with probability $p_{s,I}$:
	
	
	\begin{equation} p_{s,I}({\bf n},\theta) = \frac{\exp( \theta_{s,0,I} +  \sum_{1 \leq i \leq {\cal I}} \theta_{s,i,I} T_{s,i}({\bf n}) )}{\sum_{1 \leq s \leq N_s} \exp( \theta_{s,0,I} +  \sum_{1 \leq i \leq {\cal I}}  \theta_{s,i,I} T_{s,i}({\bf n}) )} .\end{equation}
	
	We justify the form of policy chosen, and show that the policy space contains several ``intuitively'' good policies. 
	
	We note that the action rule above is a smooth approximation to the $\max$ function. Indeed, let $x \in \RR^K$, $x^* = \maxu{k} x_k$ and $K^* = | \Set{\argmax{k} x_k} |$ the number of components of $x$ that are equal to $x^*$. Let $\gamma \in \RR$, we have that $ \frac{\exp(\gamma x_k)}{ \sum_{1 \leq k \leq K} \exp(\gamma x_k)} \tends{\gamma}{+\infty} \frac{1}{K^*}$ if $k \in \Set{\arg\max x}$ and $\frac{\exp(\gamma x_k)}{ \sum_{1 \leq k \leq K} \exp(\gamma x_k)} \tends{\gamma}{+\infty} 0$ otherwise. 
	
	We give four policies which should perform well, at least from an intuitive point of view, and give the corresponding value of $\theta$. Those policies often appear as solutions of control problems of queuing systems:
	
\begin{itemize}
\item \underline{Join the station offering the best peak rate} 

$\theta_{s,0,I} = \gamma R^{(I_s)}$ and $\theta_{s,i,I} = 0$. 

When $\gamma \to +\infty$ a user of class $(0,I)$ is attached to the base station $s^* \in \Set{ \argmax{s} R^{(I_s)}}$.
\item \underline{Join the station offering the best data rate} 

	$\theta_{s,0,I} = 0$ and $\theta_{s,i,I} = - \frac{\gamma}{ R^{(I_s)}}$. 

When $\gamma \to +\infty$ a user of class $(0,I)$ is attached to the base station $s^* \in \Set{ \argmin{s} \frac{ T_{s}({\bf n}) }{R^{(I_s)}}} = \Set{ \argmax{s} \frac{R^{(I_s)}}{ T_{s}({\bf n}) }}$.
\item \underline{Join the station with the smallest workload} 

$\theta_{s,0,I} = 0$ and $\theta_{s,i,I} = - \gamma \frac{\expec{\sigma} }{R^{(i)}}$. 

When $\gamma \to +\infty$ a user of class $(0,I)$ is attached to the base station $s^* \in \Set{ \argmin{s} \sum_{1 \leq i \leq {\cal I}} \expec{\sigma} \frac{T_{s,i}({\bf n})}{R^{(i)}}}$.

\item \underline{Join the station with the shortest queue} 

$\theta_{s,0,I} = 0$ and $\theta_{s,i,I} = - \gamma$. 

When $\gamma \to +\infty$ a user of class $(0,I)$ is attached to the base station $s^* \in \Set{ \argmin{s}  T_{s}({\bf n})}$.
\end{itemize}
	
	The existance of those four policies has two practical implications. First, finding the best parameterized policy yields performance at least as good as the previously described policies. Furthermore, if we are trying to find the optimal value of $\theta$ through an iterative search (the optimal parameterized policy), for instance using gradient descent, then the initial value of $\theta$ can be chosen as one of those four policies. This technique guarantees that even during the first iterations of the scheme, the system performance is already acceptable, as opposed to starting to a random value of $\theta$ which might yield very poor performance in the initial stages.
	
	It is noted that when the number of users in \ac{BS} $s$ is significantly larger than the other \acp{BS}, $p_{s,I}$ becomes very small and no users of $\AAb_{0}$ are served by \ac{BS} $s$. Hence, the parameterized family of policies introduces a form of load balancing.
	
\subsection{Distributed algorithm and a heuristic for improving gra\-dient estimation}\label{subsec:distributed_algo}
	Let us consider a simple setting, and make clear that the proposed algorithm can indeed be implemented in a distributed way.
	
	Each \ac{BS} $s$ has a central zone in which all users are attached to $s$, and for each neighboring \ac{BS} $s^{\prime}$ there is a zone in which users can only be attached either to $s$ or $s^{\prime}$. We will denote this zone $I_{s,s^{\prime}}$. The parameters which control the decisions for zone $I_{s,s^\prime}$ are $(\theta_{s,i,I_{s,s^\prime}}, \theta_{s^\prime,i,I_{s,s^\prime}} )_{0 \leq i \leq {\cal I}}$. It is noted that the variables $\theta_{s^{\prime\prime},i,I_{s,s^\prime}}$, $s^{\prime\prime} \notin \Set{s,s^\prime}$ will never be used, and we can simply fix $\theta_{s^{\prime\prime},i,I_{s,s^\prime}} = -\infty$ ,  $s^{\prime\prime} \notin \Set{s,s^\prime}$.

	 To alleviate notation, we will use the notation $\theta_{s,s^{\prime}}$ to denote the parameters used in the decision of association to $s$ or $s^{\prime}$. Let $\Delta_{s,s^{\prime}}$ and $z_{s,s^{\prime}}$ be the  components of $\Delta$ and $z$ respectively relative to $\theta_{s,s^{\prime}}$.
	 
	 We write once again the gradient estimation procedure equations (see \eqref{eq:zt} and \eqref{eq:Deltat}): 
\begin{align}
	z_{s,s^{\prime}}(t+1) &=\beta z_{s,s^{\prime}}(t) + \frac{\nabla_{\theta_{s,s^{\prime}}} \PP_{\theta}(a_t | s_t)}{\PP_{\theta}(a_t | s_t)} \label{eq:zt dist} \\
	\Delta_{s,s^{\prime}}(t+1) &=\Delta_{s,s^{\prime}}(t) + \frac{1}{t+1}(r_{t+1}z_{s,s^{\prime}}(t+1)-\Delta_{s,s^{\prime}}(t))	\label{eq:Deltat dist}
\end{align}

	The algorithm is distributed in the sense that actions are taken based on locally available information, namely when a user arrives in the network and could be attached to \ac{BS} $s$, then \ac{BS} $s$ needs only to know the number of active users in its neighboring \acp{BS}. Furthermore, $z_{s,s^{\prime}}(t)$ is calculated solely based on the number of active users in \ac{BS} $s$ and $s^{\prime}$.
	
	However, the calculation of $\Delta_{s,s^{\prime}}(t)$ requires to know the costs $r_{t}$, which are not a local information. For instance, if the cost is the number of active users in the whole network (case of minimization of the file transfer time as explained previously), every \ac{BS} needs to be aware of the number of active users in every \ac{BS} in the network. Another problem is that, when the number of \acp{BS} grows, the gradient estimate becomes more noisy, due to the fact that random fluctuations of the costs in all \acp{BS} will affect the estimation of the gradient with respect to $\theta_{s,s^{\prime}}$, although this parameter mainly impacts \ac{BS} $s$ and $s^{\prime}$.
 
	These two problems are serious impairments for practical applications, and we suggest a heuristic to overcome them. We assume that the cost is a sum of ``local costs'' $r = \sum_{s =1}^{N_s} r^{(s)}$, one per \ac{BS}. For instance if the cost is the total number of active users in the network, the cost of \ac{BS} $s$ is simply the number of active users in \ac{BS} $s$.

	We propose for the computation of the gradient with respect to $\theta_{s,s^\prime}$ to use only the local rewards for \ac{BS} $s$ and $s^\prime$. The proposed heuristic for gradient estimation is:	
\begin{align}
	z_{s,s^{\prime}}(t+1)&=\beta z_{s,s^{\prime}}(t) + \frac{\nabla_{\theta_{s,s^{\prime}}} \PP_{\theta}(a_t | s_t)}{\PP_{\theta}(a_t | s_t)} \label{eq:zt heur} \\
	\Delta_{s,s^{\prime}}(t+1) &=\Delta(t) \nonumber\\ &+ \frac{1}{t+1}( (r_{t+1}^{(s)} + r_{t+1}^{(s^{\prime})})z_{s,s^{\prime}}(t+1)-\Delta(t))
	\label{eq:Deltat heur}
\end{align}
	
	The heuristic is indeed fully distributed: $\Delta_{s,s^{\prime}}$ can be computed solely based on the local costs $r^{(s)}$ and $r^{(s^{\prime})}$. The intuitive explanation behind the noise reduction is that using the heuristic, any random fluctuation of the local cost in a \ac{BS} which is far away from \ac{BS} $s$ will not affect the estimation of the gradient with respect to $\theta_{s,s^{\prime}}$.
	
	We emphasize the fact that this is merely a heuristic since we cannot guarantee that the gradient estimate will be a valid ascent direction at each step. However, as shown in Section~\ref{sec:Numerical Experiments}, it performs very well numerically, and yields a considerable improvement of the gradient estimation accuracy (by a factor of $10$).
 
\section{Numerical Experiments}\label{sec:Numerical Experiments}

\subsection{Simulation setting}

\begin{figure}
\centering
\includegraphics[width =\figsize]{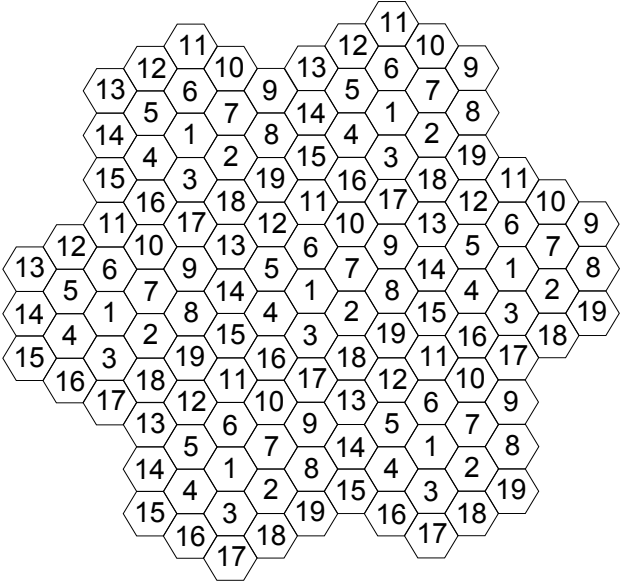}
\caption{Hexagonal network with wrap-around}
\label{fig:hexagonal_network_full}
\end{figure}

We consider a hexagonal network with $19$ \acp{BS}. In order to avoid border effects, we use a wrap-around as shown in figure \ref{fig:hexagonal_network_full}. This is essential since without the wrap-around, the stations on the outer ring would be significantly less loaded than the \acp{BS} on the inner rings, and introduce considerable bias in the simulations. As described in subsection \ref{subsec:distributed_algo} each \ac{BS} $s$ has a central zone where users are seved with a data rate of $10$ Mbps. The area of a central zone is $\frac{1}{2}$ of a cell area. For each couple of \acp{BS} $(s,s^{\prime})$, $s\neq s^\prime$, there is a zone in which users can be served by either \acp{BS} $s$ or $s^\prime$, both with a data rate of $5$ Mbps. The area of this zone is $\frac{1}{6}$ of a cell area, which is shared beween \acp{BS} $s$ and $s^{\prime}$. For the outage probability calculation, a target rate of $1$ Mbps is sought. The mean file size is $10$ Mb.

\begin{figure}
\centering
\includegraphics[width =\figsize]{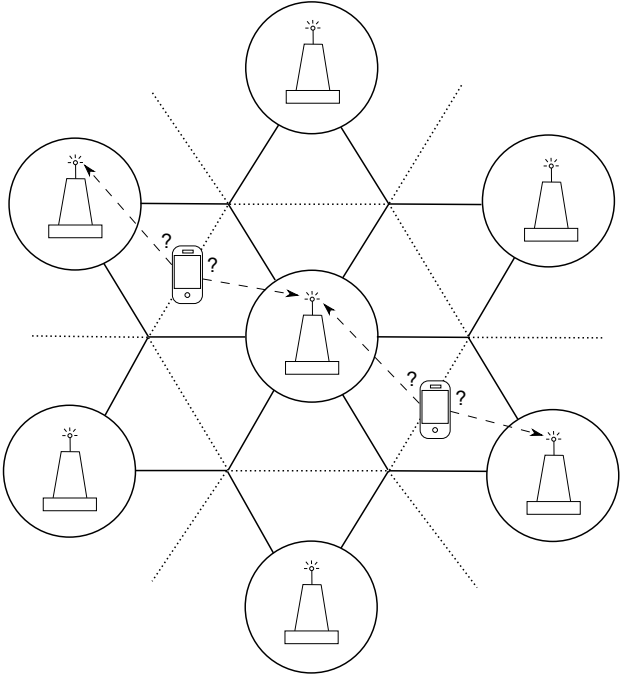}
\caption{Association problem in the simplifed setting}
\label{fig:traffic_pattern}
\end{figure}

\subsection{Results}

\begin{figure}
\centering
\includegraphics[width =\figsize]{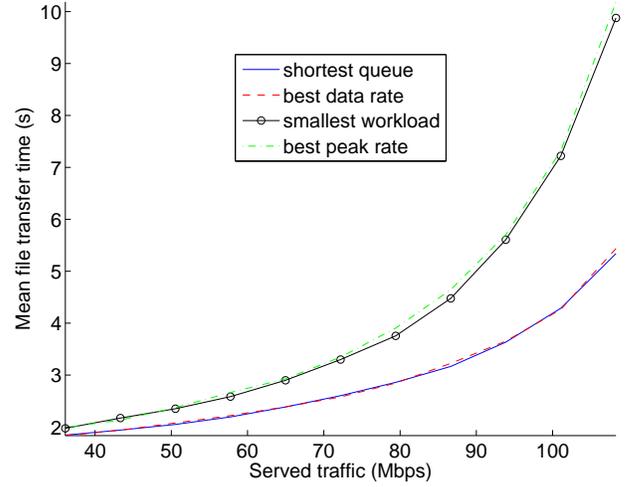}
\caption{Mean file transfer time as a function of the served traffic}
\label{fig:ftt_policies}
\end{figure}

\begin{figure}
\centering
\includegraphics[width =\figsize]{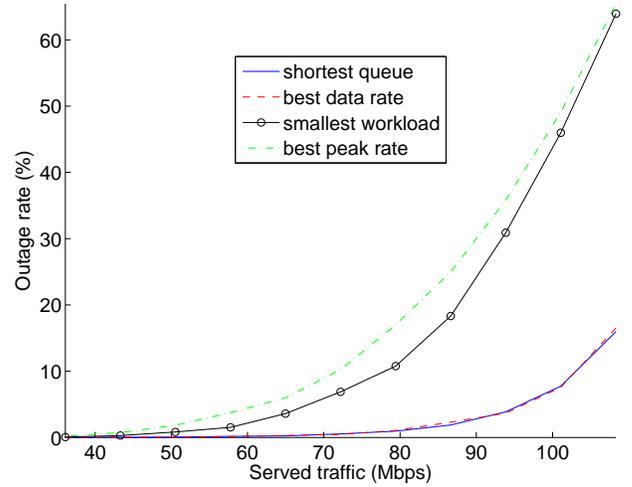}
\caption{Outage probability as a function of the served traffic}
\label{fig:outage_policies}
\end{figure}

We first compare the four policies given in the previous section.  Figures \ref{fig:ftt_policies} and \ref{fig:outage_policies} show the mean file transfer time and the outage probability for each policy, as a function of the served traffic.  The reference policy is ``best peak rate'' where users simply connect to the base station offering the best peak rate (i.e the best \ac{SINR}), without considering the loads of available \acp{BS}. The reference policy has the worst performance, since it is not load-aware, namely, it attaches users to the closest \ac{BS}, even if it is overloaded, which does not reduce network congestion when traffic is high. Policy ``smallest workload'' brings little improvement, because, even though it takes the loads into account, it can possibly admit a user when the number of active users is already large, resulting in outage. Indeed, even for a large number of active users, the workload can be small if they have almost finished their transfer, or if their data rate is high. Policies ``best data rate'' and ``shortest queue'' perform the best, and bring large improvement in both outage probability and file transfer time. For high traffic, say $100$ Mbps, policies ``best peak rate'' and ``smallest workload'' yield a mean file transfer time of $7$ s and a outage probability of $60\%$. Policies ``best data rate'' and ``shortest queue'' yield $4$ s for the mean file transfer time and $10\%$ for the outage probability. This shows that reducing congestion has a considerable impact on the network performance.
	
\begin{figure}
\centering
\includegraphics[width =\figsize]{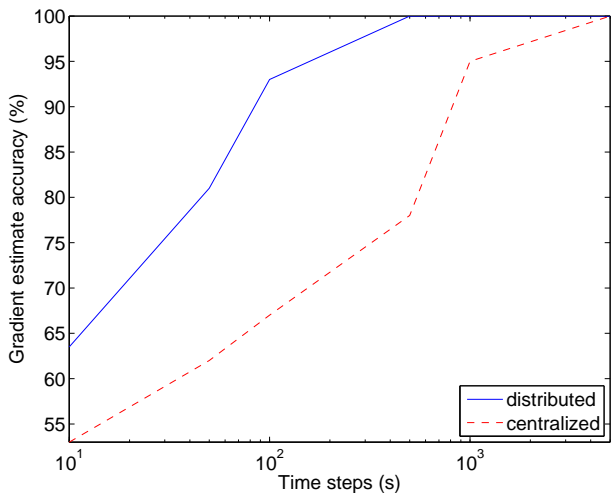}
\caption{Accuracy of gradient estimates as a function of the number of time steps}
\label{fig:policy_grad_estimate}
\end{figure}

 Figure \ref{fig:policy_grad_estimate} shows the accuracy of the gradient estimates obtained using the policy gradient method (denoted ``centralized'' in the graphs) and the proposed heuristic which allows a distributed implementation (denoted ``distributed''). We choose $\theta = 0$ for our comparison.  For a fixed number of time steps, we generate $500$ gradient estimates using both methods, and we calculate the sign of the dot product between the gradient estimate and the true gradient obtained by finite difference for a long simulation with $100,000$ time steps. If their dot product is strictly positive, then the gradient estimate is an admissible ascent direction. We plot the percentage of gradient estimates which are admissible ascent directions. The higher the percentage, the better the gradient estimate is. We can see that the accuracy of gradient estimates goes to $100\%$ when the number of time steps grows. We can also see that the proposed heuristic performs significantly better than the straighforward policy gradient. For the same level of accuracy (say $95\%$) the number of time steps required by the heuristic is $10$ times smaller than for the classical policy gradient.    

\begin{figure}
\centering
\includegraphics[width =\figsize]{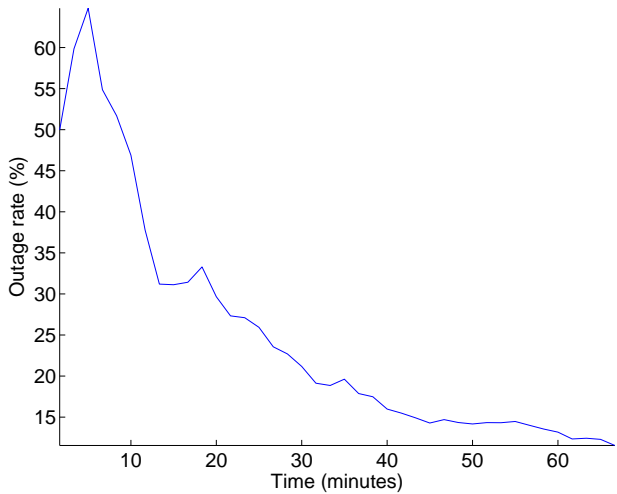}
\caption{Average cost during the learning process}
\label{fig:policy_grad_reward}
\end{figure}

\begin{figure}
\centering
\includegraphics[width =\figsize]{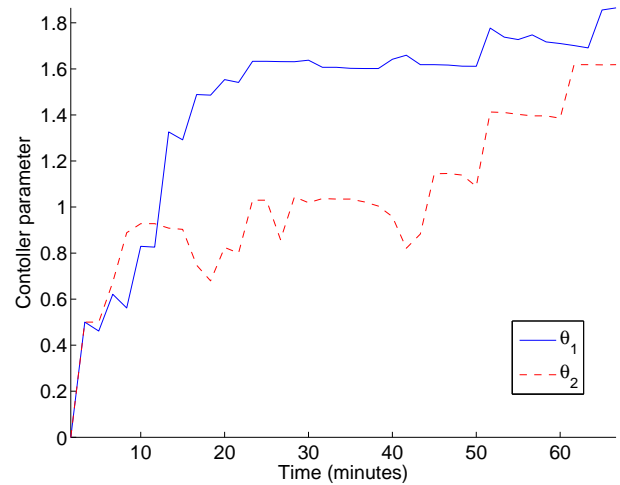}
\caption{Controller parameters during the learning process}
\label{fig:policy_grad_param}
\end{figure}

Figure \ref{fig:policy_grad_reward} and \ref{fig:policy_grad_param} show the evolution of the average cost and the corresponding controller parameter values. The total traffic is $100$ Mbps. Since the number of parameters is large ($57$ in total), only the two first components of the parameter vector are represented. For each update of $\theta$, the gradient is estimated during $100$ seconds. Starting from $\theta = 0$, and a heavily congested network, the outage probability diminishes almost monotonically. This demonstrates that the algorithm is able to find a configuration of parameters for which the congestion in the network is reduced. The algorithm convergence speed with regards to the evolution of the daily traffic is satisfactory, since in operational networks, the traffic pattern (arrival rates in each region) can reasonably be assumed fixed for at least one hour.

\section{Conclusion}\label{sec:Conclusion}
In this paper we have proposed a model for the association problem in wireless networks. This model takes into account traffic dynamics making it possible to optimize performance indicators directly perceived by users like the mean file transfer time and the outage probability. In the static framework, we show that the related association problem is tractable by classical convex optimization techniques and reduces to a routing problem. In the dynamic context, the association problem is modeled as a \ac{MDP}. An on-line Policy Gradient Reinforcement Learning method has been proposed and adapted to the association problem to optimally control the system.  A heuristic has been introduced to enable the algorithm to be implemented in a distributed manner, and improve the gradient estimation procedure dramatically. The approach has the following important advantages which make it suitable for practical implementation: 
\begin{enumerate}[(i)]
\item Its convergence to a local optimum can be proven mathematically
\item The average system performance improves monotonically, and convergence speed is compatible with typical traffic evolution in operational networks 
\item The solution is scalable, since its complexity increases linearly with the number of \acp{BS}
\item The solution can be implemented in a distributed manner.
\end{enumerate}
 Numerical experiments have demonstrated that the proposed solution performs well in practice, and effectively decreases congestion in the network.





\begin{thebibliography}{10}
\providecommand{\url}[1]{#1}
\csname url@samestyle\endcsname
\providecommand{\newblock}{\relax}
\providecommand{\bibinfo}[2]{#2}
\providecommand{\BIBentrySTDinterwordspacing}{\spaceskip=0pt\relax}
\providecommand{\BIBentryALTinterwordstretchfactor}{4}
\providecommand{\BIBentryALTinterwordspacing}{\spaceskip=\fontdimen2\font plus
\BIBentryALTinterwordstretchfactor\fontdimen3\font minus
  \fontdimen4\font\relax}
\providecommand{\BIBforeignlanguage}[2]{{%
\expandafter\ifx\csname l@#1\endcsname\relax
\typeout{** WARNING: IEEEtran.bst: No hyphenation pattern has been}%
\typeout{** loaded for the language `#1'. Using the pattern for}%
\typeout{** the default language instead.}%
\else
\language=\csname l@#1\endcsname
\fi
#2}}
\providecommand{\BIBdecl}{\relax}
\BIBdecl

\bibitem{SalahCooperativeHO}
S.~Horrich, S.~Elayoubi, and S.~Ben~Jemaa, ``On the impact of mobility and
  joint rrm policies on a cooperative wimax/hsdpa network,'' in \emph{Wireless
  Communications and Networking Conference, 2008. WCNC 2008. IEEE}, 31
  2008-april 3 2008, pp. 2027 --2032.

\bibitem{P1900.4}
S.~Buljore, H.~Harada, S.~Filin, P.~Houze, K.~Tsagkaris, O.~Holland, K.~Nolte,
  T.~Farnham, and V.~Ivanov, ``Architecture and enablers for optimized radio
  resource usage in heterogeneous wireless access networks: The ieee 1900.4
  working group,'' \emph{Communications Magazine, IEEE}, vol.~47, no.~1, pp.
  122 --129, january 2009.

\bibitem{Majed_CR_Association}
M.~Haddad, S.~Elayoubi, E.~Altman, and Z.~Altman, ``A hybrid approach for radio
  resource management in heterogeneous cognitive networks,'' \emph{Selected
  Areas in Communications, IEEE Journal on}, vol.~29, no.~4, pp. 831 --842,
  april 2011.

\bibitem{3gpp.36.300}
3GPP, ``{Evolved Universal Terrestrial Radio Access (E-UTRA) and Evolved
  Universal Terrestrial Radio Access (E-UTRAN); Overall description; Stage
  2},'' {3rd Generation Partnership Project (3GPP)}, TS {36.300}, Sep. 2008.

\bibitem{3gpp.36.902}
\BIBentryALTinterwordspacing
------, ``{Evolved Universal Terrestrial Radio Access Network (E-UTRAN);
  Self-configuring and self-optimizing network (SON) use cases and
  solutions},'' {3rd Generation Partnership Project (3GPP)}, TR {36.902}, Sep.
  2008. [Online]. Available:
  \url{http://www.3gpp.org/ftp/Specs/html-info/36902.htm}
\BIBentrySTDinterwordspacing

\bibitem{SuttonBarto}
R.~Sutton and A.~Barto, \emph{Reinforcement Learning, an Introduction}.\hskip
  1em plus 0.5em minus 0.4em\relax MIT Press, 1998.

\bibitem{PutermanMDP}
M.~L. Puterman, \emph{Markov Decision Processes: Discrete Stochastic Dynamic
  Programming}.\hskip 1em plus 0.5em minus 0.4em\relax Wiley-Interscience,
  2005.

\bibitem{WilliamsReinforce}
\BIBentryALTinterwordspacing
R.~J. Williams, ``Simple statistical gradient-following algorithms for
  connectionist reinforcement learning,'' \emph{Machine Learning}, vol.~8, pp.
  229--256, 1992, 10.1007/BF00992696. [Online]. Available:
  \url{http://dx.doi.org/10.1007/BF00992696}
\BIBentrySTDinterwordspacing

\bibitem{BaxterBartlettPolicyGradient1}
\BIBentryALTinterwordspacing
J.~Baxter and P.~L. Bartlett, ``{Infinite-Horizon Policy-Gradient
  Estimation},'' \emph{Journal of Artificial Intelligence Research}, vol.~15,
  pp. 319--350, 2001. [Online]. Available:
  \url{http://citeseerx.ist.psu.edu/viewdoc/summary?doi=10.1.1.21.8723}
\BIBentrySTDinterwordspacing

\bibitem{BaxterBartlettPolicyGradient2}
\BIBentryALTinterwordspacing
J.~Baxter, P.~L. Bartlett, and L.~Weaver, ``{Experiments with Infinite-Horizon
  Policy-Gradient Estimation},'' \emph{Journal of Artificial Intelligence
  Research}, vol.~15, pp. 351--381, 2001. [Online]. Available:
  \url{http://www.jair.org/papers/paper807.html}
\BIBentrySTDinterwordspacing

\bibitem{CombesPEVA2011}
R.~Combes, Z.~Altman, and E.~Altman, ``Scheduling gain for frequency-selective
  rayleigh-fading channels with application to self-organizing packet
  scheduling,'' \emph{Performance Evaluation}, Feb. 2011.

\bibitem{CombesCoverageCapacityMIMO2011}
------, ``A self-optimization method for coverage-capacity optimization in
  ofdma networks with mimo,'' in \emph{Value Tools 2011}.

\bibitem{CombesWiopt2011}
R.~Combes, S.~E. Elayoubi, and Z.~Altman, ``Cross-layer analysis of scheduling
  gains: Application to lmmse receivers in frequency-selective rayleigh-fading
  channels,'' in \emph{WiOpt 2011}, 2011.

\bibitem{Bonald-dimensioning}
T.~Bonald and A.~Prouti\`ere, ``Wireless downlink data channels: User
  performance and cell dimensioning,'' in \emph{ACM Mobicom}, 2003.

\bibitem{little}
\BIBentryALTinterwordspacing
J.~D.~C. Little, ``{A Proof for the Queuing Formula: L= $\lambda$ W},''
  \emph{Operations Research}, vol.~9, no.~3, pp. 383--387, 1961. [Online].
  Available: \url{http://dx.doi.org/10.2307/167570}
\BIBentrySTDinterwordspacing

\bibitem{ConvexOptim}
S.~Boyd and L.~Vandenberghe, \emph{Convex Optimization}.\hskip 1em plus 0.5em
  minus 0.4em\relax Cambridge University Press, 2004.

\bibitem{Blackwell}
D.~Blackwell, ``Discrete dynamic programming,'' \emph{Annals of Mathematical
  Statistics}, vol.~33, no.~2, pp. 719--726, 1962.

\end{thebibliography}
\end{document}